# On Mobile DNA in Artificial Regulatory Networks: Evolving Functional and Structural Dynamism

Larry Bull

Department of Computer Science & Creative Technologies

University of the West of England

Bristol BS16 1QY, U.K.

+44 (0)117 3283161

Larry.Bull@uwe.ac.uk

Abstract

There is a growing body of work considering the use of representations based upon genetic regulatory networks. This paper uses a recently presented abstract, tunable Boolean regulatory network model to explore aspects of mobile DNA, such as transposons, within these dynamical systems. The significant role of mobile DNA in the evolution of natural systems is becoming increasingly clear. Whilst operators loosely based upon transposons have previously been used within evolutionary computation, their use within regulatory network representations enables the potential exploitation of numerous new mechanisms. This paper shows how dynamically controlling network node connectivity and function via transposon-inspired mechanisms can be selected for under non-stationary and coevolutionary scenarios, including when such changes are heritable.

1. Introduction

A number of mobile DNA mechanisms exist through which changes in genomic structure can occur in ways other than copy errors, particularly via transposable elements (e.g., see [Craig et al., 2002] for an overview). Mobile genetic elements such as transposons are DNA sequences that may be either copied or removed and then inserted at a new position in the genome [McClintock, 1987]: retrotransposons use an intermediary RNA copy of themselves for "copying and pasting", whereas DNA transposons rely upon specific proteins for their "cutting and pasting" into new sites, respectively. The targeting of a new position ranges from the very specific, typically by exploiting sequence recognition proteins, to more or less arbitrary movement. These processes, insertion in particular, are often reliant upon proteins produced elsewhere within the genome. Transposons are found widely in both prokaryotes and eukaryotes, and they have been associated with many significant evolutionary innovations (e.g., see [Kazazian, 2004] for an overview).

Transposable elements can therefore change the behaviour of a given cell: insertion into a gene will typically disrupt its coding sequence, i.e., it will be mutated, insertion next to a gene may affect its subsequent regulation, e.g., the mobile element's regulatory sequence may take control of the gene, the act of excision can leave behind DNA fragments which cause a change in the sequence at that location, coding segments between transposons can be moved with them, etc. The effects of such movement can be beneficial or detrimental to a cell. Perhaps the most significant aspect of transposons is that these effects occur *during* the organism/cell's lifetime. That is, *such structural changes are made to a genome based upon the actions of its own regulatory processes in response to its internal and external environment*. Moreover, such changes can be inherited. Thus, as has recently been highlighted [Shapiro, 2011], genomes should be viewed as read-write systems with embedded change/search heuristics.

In all known prior uses of transposon-inspired mechanisms within evolutionary computation their role has only been considered at the point of reproduction, being viewed as a form of recombination (e.g., [Simoes & Costa, 1999]) or duplication (e.g., [Ferreira, 2001]). That is, transposons have *not* been considered as the underlying mechanism for an on-going, context dependent restructuring process as seen in nature, which is possible within artificial genetic regulatory network representations. This paper extends a recent initial study which began consideration of the dynamic role of mobile DNA within regulatory network representations [Bull, 2012a]. In particular, an aspect of transposable elements within a genetic regulatory network (GRN) was explored using an extension of a well-known, simple GRN formalism – random Boolean networks (RBN) [Kauffman, 1969]. A general overview of RBN can be in [Gershenson, 2002] and a review of their use as a dynamical representation scheme for evolutionary computation can be found in [Bull, 2012b].

With the aim of enabling the systematic exploration of GRN as a general representation scheme, a simple approach to combining them with abstract fitness landscapes has recently been presented [Bull, 2012b]. More specifically, RBN were combined with the NK model of fitness landscapes [Kauffman & Levin, 1987], not least since much work in evolutionary computation has utilized the tunable model (e.g., after [Altenberg, 1994]). In the combined form – termed the RBNK model [Bull, 2012b] – a simple relationship between the states of $N$ randomly assigned nodes within an RBN is assumed such that their value is used within a given NK fitness landscape of trait dependencies. The approach was also extended to enable consideration of coevolutionary optimization using the related NKCS landscapes [Kaufmann & Johnsen, 1991] – the RBNKCS model [Bull, 2012b].

This paper explores the potential for dynamic node connectivity and function based upon the current state of the GRN and its environment during evaluation in both the RBNK and RBNKCS models.

2. The RBNK Model

Within the traditional form of RBN, a network of $R$ nodes, each with a randomly assigned Boolean update function and $B$ directed connections randomly assigned from other nodes in the network, all update synchronously based upon the current state of those $B$ nodes. Hence those $B$ nodes are seen to have a regulatory effect upon the given node, specified by the given Boolean function attributed to it. Since they have a finite number of possible states and they are deterministic, such networks eventually fall into an attractor. It is well-established that the value of $B$ affects the emergent behaviour of RBN wherein attractors typically contain an increasing number of states with increasing $B$ (see [Kauffman, 1993] for an overview). Three phases of behaviour exist: ordered when $B=1$, with attractors consisting of one or a few states; chaotic when $B \geq 3$, with a very large number of states per attractor; and, a critical regime around $B=2$, where similar states lie on trajectories that tend to neither diverge nor converge (see [Derrida & Pomeau, 1986] for formal analysis).

As shown in Figure 1, in the RBNK model $N$ nodes in the RBN are chosen as "outputs", i.e., their state determines fitness using the NK model. Kauffman and Levin [1987] introduced the NK model to allow the systematic study of various aspects of fitness landscapes (see [Kauffman, 1993] for an overview). In the standard model an individual is represented by a set of $N$ (binary) genes or traits, each of which depends upon its own value and that of $K$ randomly chosen others in the individual. Thus increasing $K$, with respect to $N$, increases the epistasis. This increases the ruggedness of the fitness landscapes by increasing the number of fitness peaks.

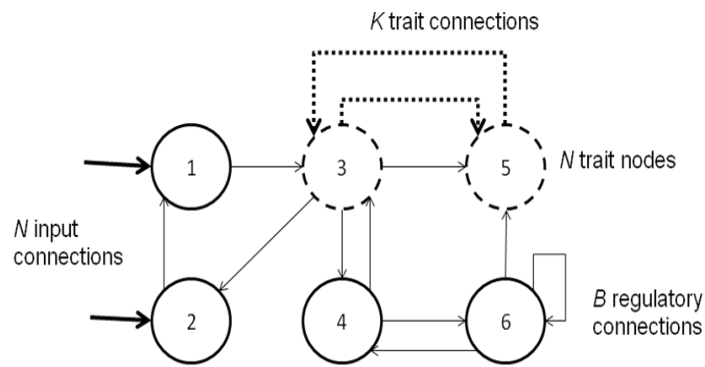

An $R$=6, $B$=2, $N$=2, $K$=1 network with two inputs

Fig 1: Variant of the RBNK model used with $N$ external inputs applied thereby making the GRN sensitive to its environment.

The NK model assumes all epistatic interactions are so complex that it is only appropriate to assign (uniform) random values to their effects on fitness. Therefore for each of the possible $K$ interactions, a table of $2^{(K+1)}$ fitnesses is created, with all entries in the range 0.0 to 1.0, such that there is one fitness value for each combination of traits. The fitness contribution of each trait is found from its individual table. These fitnesses are then summed and normalised by $N$ to give the selective fitness of the individual. Exhaustive search of NK landscapes [Smith & Smith, 1999] suggests three general classes exist: unimodal when $K$=0; uncorrelated, multi-peaked when $K$>3; and, a critical regime around $0<K<4$, where multiple peaks are correlated.

The combination of the RBN and NK model enables a systematic exploration of the relationship between phenotypic traits and the genetic regulatory network by which they are produced. It was previously shown how achievable fitness decreases within increasing $B$, how increasing $N$ with respect to $R$ decreases achievable fitness, and how $R$ can be decreased without detriment to achievable fitness for low $B$ [Bull, 2012b]. In this paper, following [Bull, 2012a], a simple scheme is adopted: $N$ phenotypic traits are attributed to arbitrarily chosen nodes within the network of $R$ genetic loci and with environmental inputs applied to the first $N$ loci (Figure 1). Hence the NK element creates a tuneable component to the overall fitness landscape with behaviour (potentially) influenced by the environment. In this paper, the inputs are $N$-bit random strings per NK landscape.

## 3. Transposable Elements in the RBNK Model

4.1 Structural Dynamism

In the aforementioned initial study of mobile DNA in RBN [Bull, 2012a], structural dynamism was seen as a consequence of the actions of DNA transposons. That is, the cutting-and-pasting of segments of DNA was seen as causing a change in the connectivity structure of the GRN. Here nodes were extended to (potentially) include a second set of $B'$ connections to defined nodes. Each such dynamic node also performed an assigned rewiring function based upon the current state of the $B'$ nodes, as shown in Figure 2. Hence on each cycle, each node updates its state based upon the current state of the $B$ nodes it is connected to using the Boolean logic function assigned to it in the standard way. Then, if that node is also structurally dynamic, those $B$ connections are altered according to the current state of the $B'$ nodes it is connected to using its rewiring table. The moving of the $B$ connections of a given node via the actions/states of the $B'$ nodes is therefore seen as an abstraction of one or more of the possible effects of a mobile element as discussed above, triggered by one or more of the $B'$ nodes, causing a change in the regulatory network which affects the given node. For simplicity, the number of regulatory connections ($B$) is assumed to be the same as for rewiring ($B'$).

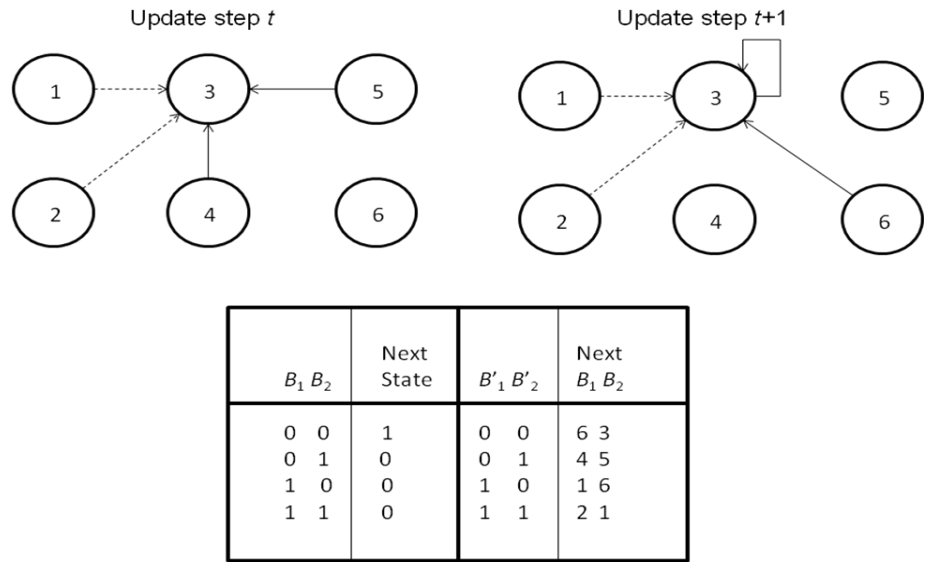

Fig 2: Example RBN with structural dynamism as in [Bull, 2012a]. The look-up table and connections for node 3 are shown in an *R*=6, *B*=2 network. Nodes capable of rewiring have *B'* extra structure regulation connections into the network (dashed arrows) and use the state of those nodes to alter the standard *B* transcription regulation connections (solid arrows) on the next update cycle (*B'*=2). Thus in the RBN shown, node 3 is a dynamic node and uses nodes 1 and 2 to determine any structural changes. At update step *t*, node 3 is shown using the states of nodes 4 and 5 to determine its state for the next cycle. Assuming all nodes are at state '0', the given node above would transit to state '1' for the next cycle and source its *B* inputs from nodes 6 and 3 on that subsequent cycle, as defined in the first row of the table shown. A DNA transposon-like mediated change in the regulation network is said to have occurred and the genome rewritten – the *B* source connection ids are altered.

For simplicity with respect to the underlying evolutionary search process, a genetic hillclimber was considered in [Bull, 2012a], as it is here. Each RBN is represented as a list to define each node's start state, Boolean function, *B* connection ids, *B'* connection ids, connection changes table entries, and whether it is a dynamic node or not. Mutation can therefore either (with equal probability): alter the Boolean function of a randomly chosen node; alter a randomly chosen *B* connection (used as the initial connectivity if a dynamic node); alter a node start state; turn a node into or out of being a dynamic rewiring node; alter one of the rewiring entries in the look-up table if it is a dynamic node; or, alter a randomly chosen *B'* connection, again only if it is a dynamic node. A single fitness evaluation of a given GRN is ascertained by updating each node for 100 cycles from the genome defined start states. At each update cycle, the value of each of the *N* trait nodes in the GRN is used to calculate fitness on the given NK landscape. The final fitness assigned to the GRN is the average over 100 such updates here. A mutated GRN becomes the parent for the next generation if its fitness is higher than that of the original. In the case of fitness ties the number of dynamic nodes is considered, with the smaller number favoured, the decision being arbitrary upon a further tie. Hence there is a slight selective pressure against structural dynamism. Here $R=100$, $N=10$ and results are averaged over 100 runs - 10 runs on each of 10 landscapes per parameter configuration - for 30,000 generations. As in [Bull, 2012b], $0<B\leq5$ and $0\leq K<5$ are used.

It was shown that for $B<3$, starting with no dynamic nodes, around 5% of nodes became dynamic in a non-stationary environment but that dynamic nodes were not incorporated in a stationary environment [Bull, 2012a]. The non-stationary case was created by applying a second input string and use of a second fitness landscape for the latter half of an evaluation. Analysis of the rewiring behaviour in the low *B* cases showed that the dynamic nodes typically fired for *only* the first few update cycles after both initialization and the switch in input halfway through the lifecycle. Thus it appears that in low connectivity networks *the evolutionary process was exploiting structural dynamism to help shape the attractor space of the RBN such that high fitness was reliably reached depending upon the environmental input and GRN state*. This non-stationary case was somewhat motivated by the growing number of examples of environmentally triggered – typically under stress conditions - genomic rearrangements found in a wide variety of organisms (e.g., see [Shapiro, 2011]).

As briefly noted at the end of [Bull, 2012a], the structural dynamism can be defined in a position relative way instead of the explicit addressing form shown in Figure 2. That is, the entries in the columns for the *B'* nodes become movements in a range relative to their current connection id's, e.g., using a range +/-5 nodes, stopping at either end of the genome. Results were largely unchanged using this approach, one which may be seen as more akin to a DNA transposon excising itself at a given location and inserting itself at the next available appropriate position along the genome – which might be viewed as being a less specific insertion process than that shown in Figure 2. Whilst not previously reported, analysis of the underlying behaviour indicates that the dynamic nodes typically experience constant rewiring during execution, usually moving between a finite set of connections as the RBN moves through its (deterministic) attractor. That is, the rewiring connections were typically made to nodes which alter their state within the attractor of the network, whereas with the previous mechanism the rewiring connections were typically made to nodes which do not alter their state within the resulting attractor.

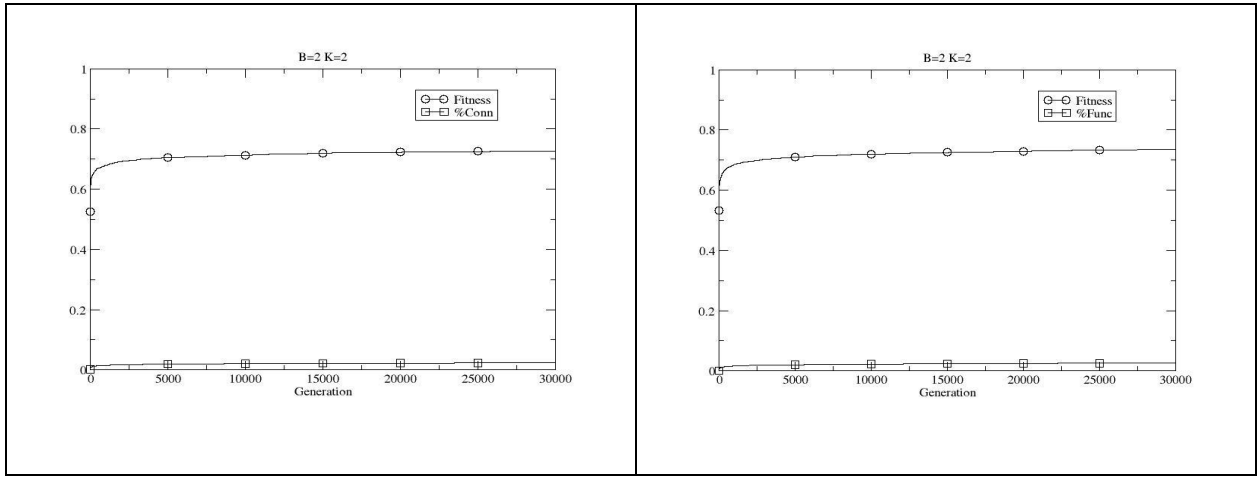

Fig 3: Examples of structural (left) and functional (right) dynamism.

Figure 3 (left) shows an example result of the position relative structural dynamism representation in a non-stationary environment. Here, unlike in the previous study, the fitness landscape does not deterministically switch between the same two per evaluation lifecycle, i.e., an RBN does not experience landscape 1 for 50 cycles and then landscape 2 for 50 cycles. Rather, for the first 50 cycles an offspring experiences the same input-landscape combination as its parent did on its last 50 cycles, before experiencing a new, random input-landscape combination for its last 50 cycles. Hence the GRN must evolve to exist in an environment of constant, significant novelty. As can be seen, the percentage of dynamic nodes stabilizes at around 2% on average.

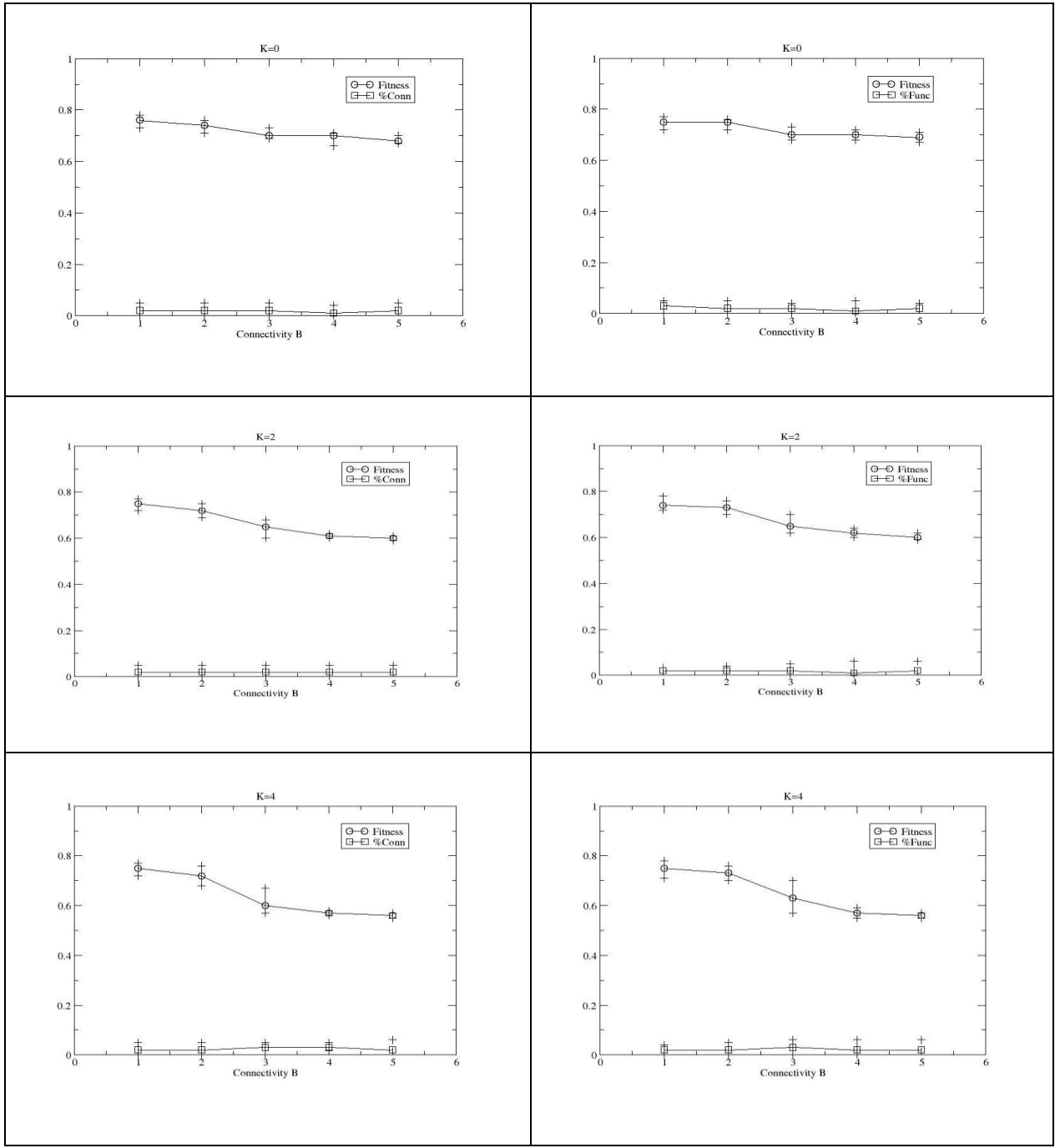

Fig 4: Performance of the structurally dynamic (left column) and functionally dynamic (right column) RBN, after 30,000 generations. Error bars show min and max values in all graphs.

Figure 4 (left column) shows further examples of how this result holds for all *B*, regardless of the underlying topology of the fitness landscapes. Fitness is typically significantly decreased (T-test, p<0.05) for *B*>2, i.e., with traditionally chaotic networks, as previously reported [Bull, 2012b]. The same general results were found for varying the size of the networks, e.g., *R*=200, the final percentage of dynamic nodes tends to vary proportionally, i.e., ~5%, but not with consistent statistical significance (not shown). Given that both the number and size of attractors are known to be proportional to *R* [Kauffman, 1993] this is perhaps to be expected. Dynamic nodes were not selected for in the stationary, single input-landscape case, as expected (not shown). There is no significant difference in performance compared to the previous, more specific movement mechanism (not shown).

3.2 Functional Dynamism

Probabilistic RBN (e.g., see [Shmulevich & Dougherty, 2010]) allow for a change in node function within a given set according to a fixed distribution. It has long been noted (e.g., see [Kauffman, 1984]) that a bias in the Boolean function space of the traditional RBN - that is, a deviation from the expected average probability *P* of 0.5 for either state as the output - reduces the number of attractors and their size for a given number of nodes and connectivity. Following the node relative adjustment scheme used for connectivity, a deterministic context-sensitive form of dynamic node can be defined which incrementally alters the number of 0's or 1's in the Boolean function table for that node, as shown in Figure 5. Hence on each cycle, each node updates its state based upon the current state of the *B* nodes it is connected to using the Boolean logic function assigned to it in the standard way. Then, if that node is also functionally dynamic, the node function is altered according to the current state of the *B*' nodes it is connected to. Entries in the *B*' columns can now be either a 0 or 1. A node's Boolean logic function is stored as a binary string of $2^B$ bits. The first bit in that logic function table which is not the same as the entry in the dynamic table indexed by the current state of the *B*' connections is flipped. In this way node function can be varied in an incremental way based upon the current internal and external state of the RBN, here seen as capturing different aspects of mobile DNA than the structural dynamism.

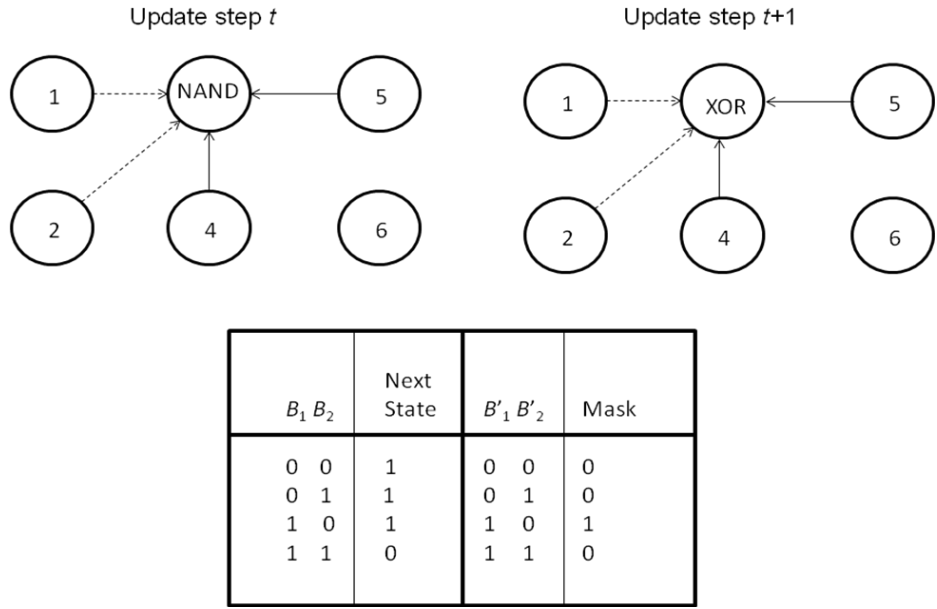

Fig 5: Example RBN with functional dynamism. The look-up table and connections for node 3 are shown in an *R*=6, *B*=2 network. Nodes capable of re-functioning have *B'* extra regulation connections into the network (dashed arrows) and use the state of those nodes to alter the node's Boolean logic function on the next update cycle (*B'*=*B*). Thus in the RBN shown, node 3 is a dynamic node and uses nodes 1 and 2 to determine any functional changes. At update step *t*, node 3 is shown using a two-bit NAND function to determine its state for the next cycle (encoded as 1110). Assuming all nodes are at state '0', the given node above would transit to state '1' for the next cycle and alter the first non-zero bit in its function table on that subsequent cycle, as defined in the first row of the table shown, hence changing to XOR (encoded as 0110). A DNA transposon-like mediated change in the regulation network is said to have occurred and the genome rewritten.

Figure 3 (right) shows an example result of the incremental functional dynamism representation in a non-stationary environment, as used above. As can be seen, the percentage of dynamic nodes again stabilizes at around 2% on average. Figure 4 (right column) shows the typical behaviour for functional dynamism is the same as for structural dynamism (left column) and analysis of the re-functioning indicates similar constant cyclic adjustment within an attractor.

4.3 Combined Dynamism

The previous two transposon-inspired mechanisms can be combined such that dynamic nodes may be either structurally or functionally variable (in principle, a node could be dynamic in both ways but that has not been explored here). Figure 6 (left column) shows the behaviour of RBN with the combined dynamism. As can be seen, dynamic nodes typically constitute around 3% of the networks, with analysis indicating both types of dynamism are selected for in roughly equal proportion.

In the above experiments an offspring's nodes were initialized according to their genome specification as in the previous work on evolving RBN, regardless of the final connectivity pattern and/or node functionality of the parent due to the effects of any dynamic nodes it contained. That is, genomic rearrangements were not inherited. However, it has long been argued that transposon-mediated changes are a principle source of heritable variation in natural evolution (e.g., [Shapiro, 1992]). Figure 6 (right column) shows the evolutionary behaviour of the dynamic RBN when the parent's final network structure, functionality and node states are inherited by the offspring. The very first RBN are assigned random connectivity, functions and node start states. The results indicate there is no significant change in fitness or the percentage of dynamic nodes (T-test, $p \geq 0.05$) to the previous version for all $B$ and $K$ combinations used (compare to left column). That is, there is no significant change in the evolutionary process when the affects of the transposable elements are inheritable.

The aforementioned initial study of transposons in RBN [Bull, 2012a] explored this aspect such that the entries in the look-up tables for the rewiring of $B$ connections in any dynamic nodes and their $B'$ connections were re-assigned arbitrarily in offspring, i.e., not inherited from the parent. Fitness was typically significantly decreased when $B<3$ and $K<4$. This again reinforces the view that evolution was able to exploit rewiring for low connectivity networks to shape the attractor space, with the caveat it is most beneficial whilst the underlying fitness landscape is largely correlated (see Section 3); when the positions of fitness optima are structured, the rewiring is able to embed a simple heuristic to change GRN behaviour to suit which of the two fitness landscapes it finds itself upon. This has been explored here and, perhaps as expected, due to the purely random nature of the changes in the fitness landscapes, there is no statistically significant difference in fitness when dynamic nodes are randomly re-defined in offspring (T-test, $p \geq 0.05$, not shown). However, dynamic nodes of both types were still selected for indicating their utility for adjusting the attractor space of the RBN in (randomly changing) non-stationary fitness landscapes.

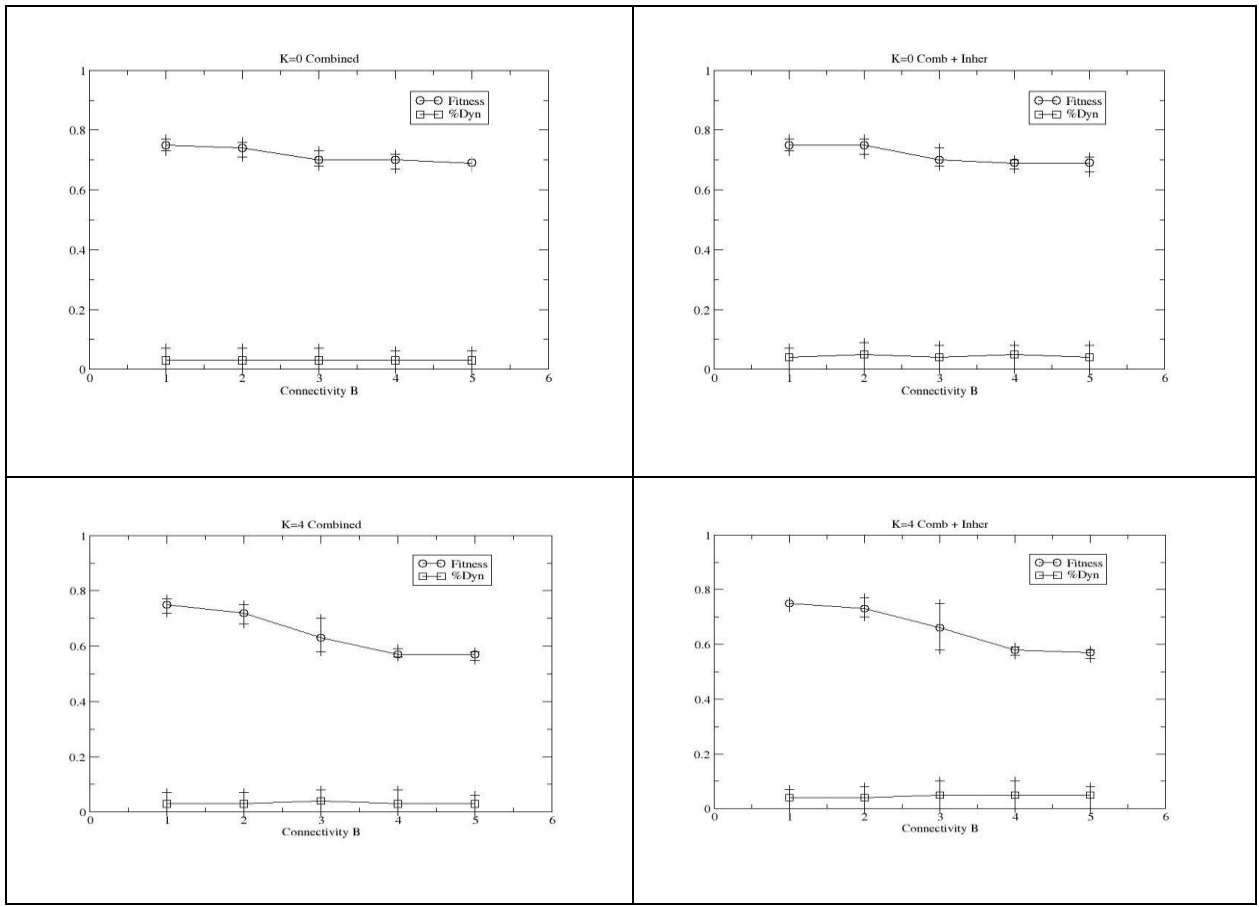

Fig 6: Performance of the fully dynamic RBN where offspring do not (left column) and do (right column) inherit the genomic changes made during the lifecycle of the parent, after 30,000 generations.

## 4. Transposable Elements in the RBNKCS Model

In the preceding experiments the GRN were coupled to external inputs and experienced changes in their environment in a relatively simplistic way. As noted elsewhere [Bull, 2012b], of particular interest is work which includes multiple cells, i.e., as in the natural case, those where the activity of the GRN primarily affect, and are affected by, other GRN (see [Bull, 2012b] for an overview).

Kauffman and Johnsen [1991] presented a coevolutionary variant of the NK model – the NKCS model. Here each node/gene is coupled to $K$ others locally and to $C$ (also randomly chosen) within each of the $S$ other species/individuals with which it interacts or is dependent upon in some way. It is shown that as $C$ increases, mean performance drops and the time taken to reach an equilibrium point increases, along with an associated decrease in the equilibrium fitness level. That is, adaptive moves made by one partner deform the fitness landscape of its partner(s), with increasing effect for increasing $C$. As in the NK model, it is again assumed all intergenome ($C$) and intragenome ($K$) interactions are so complex that it is only appropriate to assign random values to their effects on fitness. Therefore for each of the possible $K+C$x$S$ interactions, a table of $2^{(K+1+CxS)}$ fitnesses is created, with all entries in the range 0.0 to 1.0, such that there is one fitness value for each combination of traits. The fitness contribution of each gene is found from its individual table. These fitnesses are then summed and normalised by $N$ to give the selective fitness of the total genome (see [Kauffman, 1993] for an overview).

The RBNK model is easily extended to consider the interaction between multiple GRN based on the NKCS model – the RBNKCS model [Bull, 2012b]. As Figure 7 shows, it is here assumed that the current state of the $N$ trait nodes of one network provide input to a set of $N$ internal nodes in each of its coupled partners, i.e., each serving as one of their $B$ connections. Similarly, the fitness contribution of the $N$ trait nodes considers not only the $K$ local connections but also the $C$ connections to its $S$ coupled partners' trait nodes.

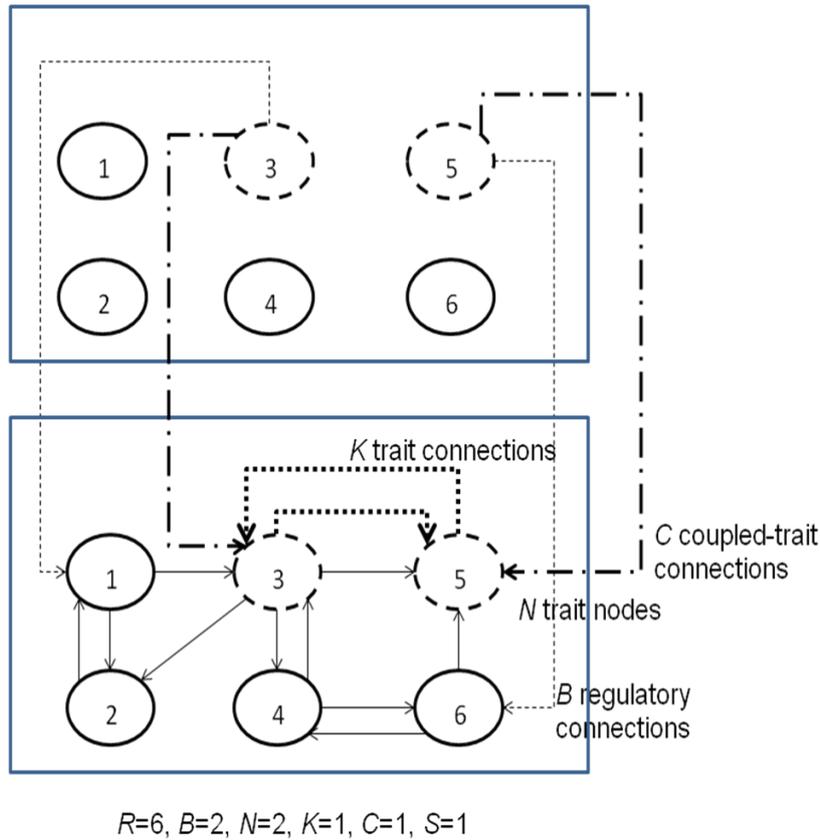

Fig. 7: Example RBNKCS model. Connections for only one of the two coupled networks are shown, without any dynamic nodes for clarity.

5.1 Multicellularity

The role of transposons in multicellular natural systems is only just beginning to be explored. There is some evidence that retrotransposons are active in embryogenesis to create somatic variation (e.g., [Kano et al., 2009]), although the effects are often linked to disease. Most significantly, it is clear that retrotransposons create diversity during neurogenesis within human brain areas such as the hippocampus (e.g., [Coufal et al., 2009]).

The case of two interacting cells has previously been explored with the *RBNKCS* model, where one is the daughter (clone) of the other, i.e., *S*=1 [Bull, 2012b]. In all of the following experiments both structural and functional dynamism is included into this model, along with the inheritance of changes, i.e., as in the latter experiments in Section 4 (Figure 6, right column). Figure 8 shows example results of the (reproducing) mother cell for various *B, K*, and *C* where the mother and daughter cells exist on the same NKCS landscape. All other details were as in Section 4.3 and $0<C\leq5$. The mother cell is updated through one cycle and then both update in turn for 100 cycles, thereby introducing some asymmetry in GRN states into the model, with the mother receiving the average fitness of the two cells. For $B<3$, i.e., the typically high fitness cases, dynamism is not selected for on average, regardless of the value of *C*. Recall such networks typically exhibit a point or small attractor and hence the GRN of the two cells exist in relatively static environments. Dynamism is selected for in the high *B* cases, as in most results thus far. However, it is selected for to previously unseen levels, typically ~20%, and with the relative difference in fitness to the low *B* cases greatly reduced for low inter-cell coupling, particularly when the underlying fitness landscape is correlated, i.e. $K<4$. This effect disappears when the degree of coupling between the two cells increases. That is, for low *C* evolution appears able to exploit dynamism more effectively for high *B* than in any previous case considered above. Here the environment of each RBN is a copy of itself. Thus it appears dynamism of both kinds is used equally to alter the attractor space of the two cells such that their inherently chaotic behaviour (Section 2) is mutually contained and hence relatively high fitness levels are achieved, i.e., cooperative behavior emerges.

Figure 9 shows results where some form of differentiation in the daughter's role is included such that it exists on a different NKCS landscape to the mother. The results indicate that for $B=1$, regardless of *K* and *C*, dynamism is selected for in around 1% of nodes on average. For $B=2$ there is on average around 4% of dynamic nodes per GRN over various *K* and *C*. This is a statistically significant increase (T-test, $p<0.05$) in dynamism compared to the equivalent homogeneous case above. Again, there are roughly equal amounts of structural and functional dynamism. The ability of evolution to exploit dynamism in the high *B* and low *C* cases typically remains.

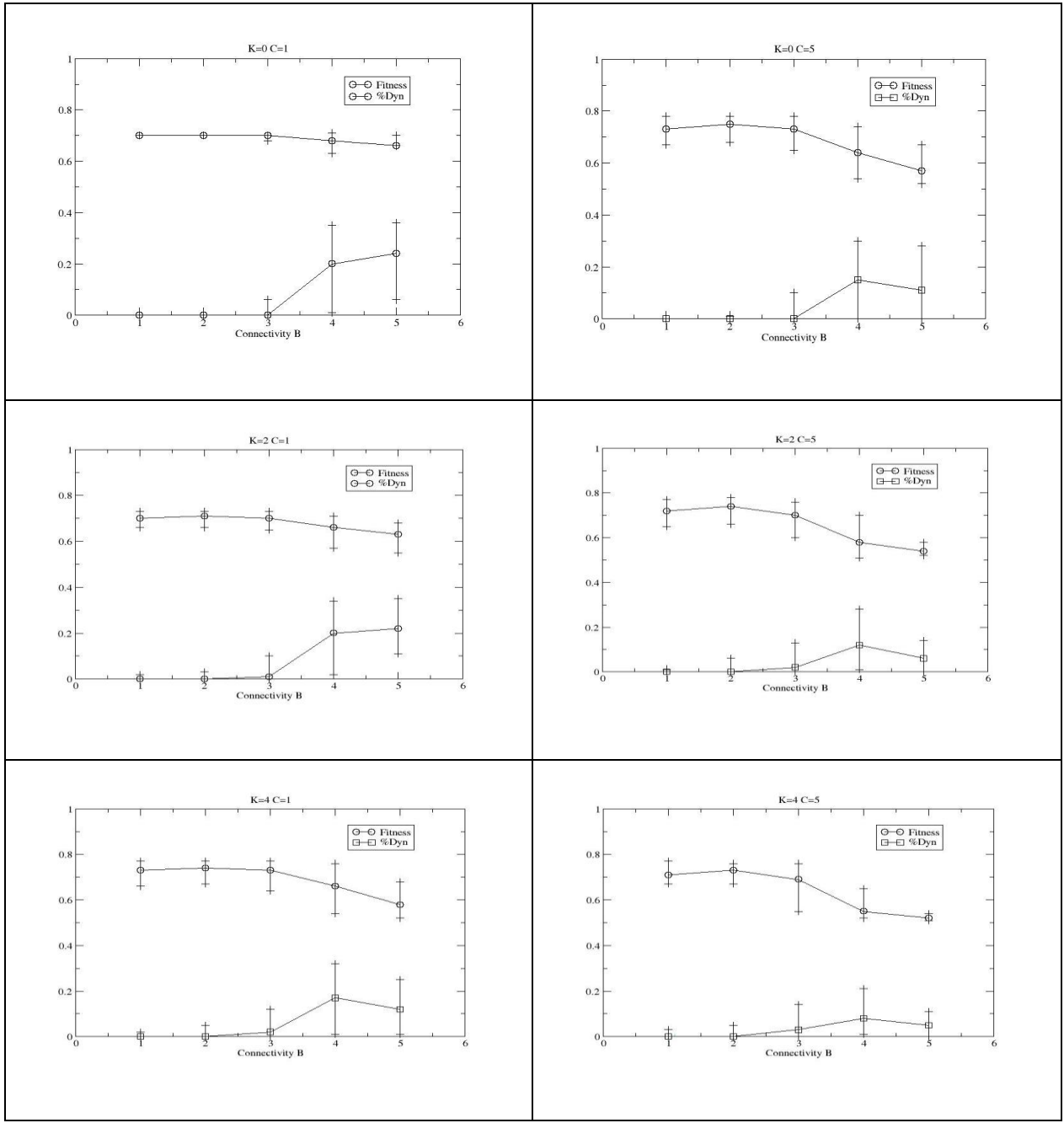

Fig 8: Performance of the fully dynamic RBN as a homogeneous, two-celled multicellular organism, after 30,000 generations.

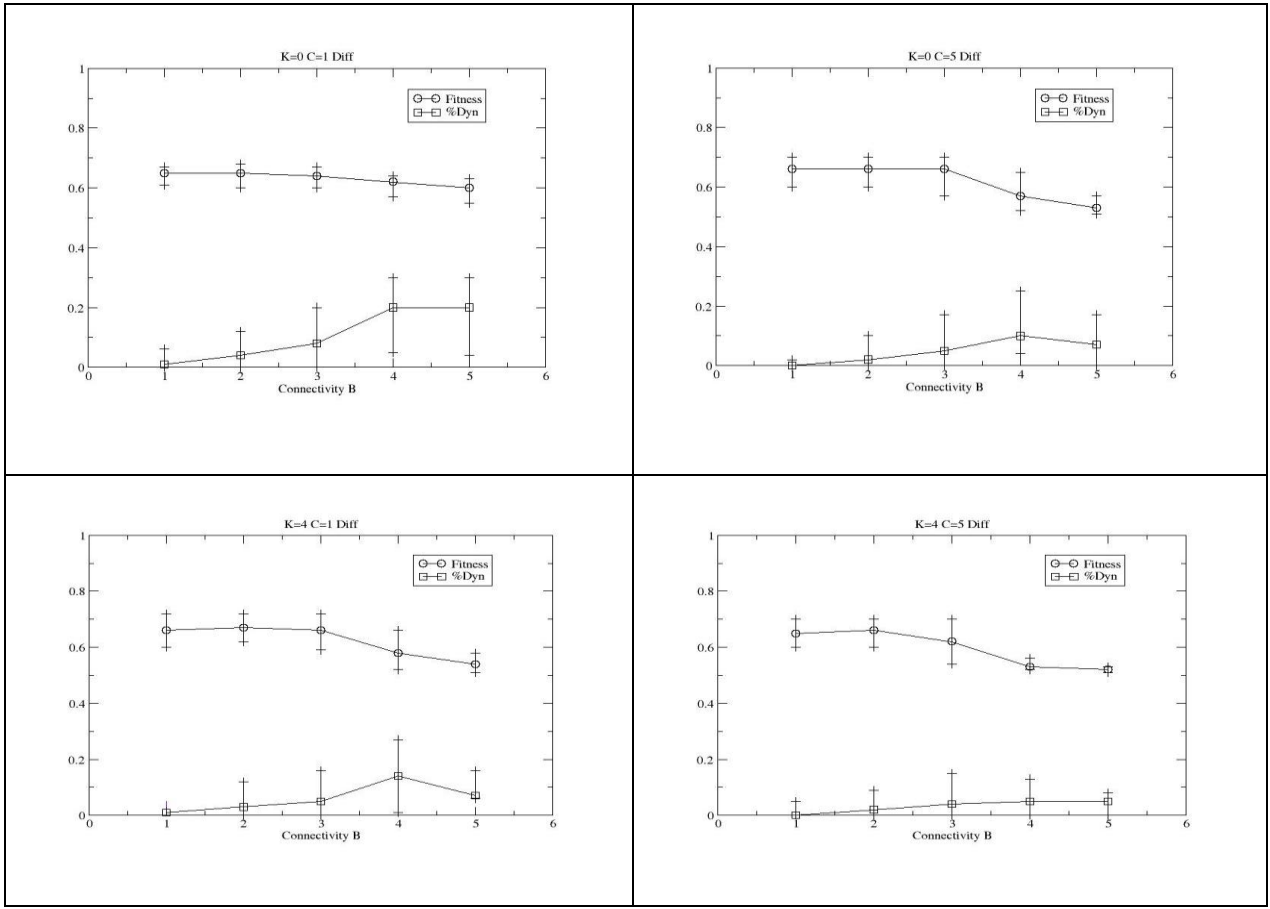

Fig 9: Performance of the fully dynamic RBN as a heterogeneous, two-celled multicellular organism, after 30,000 generations.

4.2 Coevolution

The case of two coevolving GRN has previously been explored using the RBNKCS model, each evolved separately on their own NKCS fitness landscape for their $N$ external traits [Bull, 2012b]. Each network again updates in turn for 100 cycles, as above. The fitness of one network is then ascertained and an evolutionary generation for that network is undertaken. The mutated network is evaluated with the same partner – structurally and functionally reset - as the original and it becomes the parent under the same criteria as used above. Then the second population network is evaluated with that network, before a mutated form is created and evaluated against the same partner. Again, it is adopted if fitter or has fewer dynamic nodes. One generation is said to have occurred when all four steps have been undertaken. Due to the symmetry of the simulations, only the fitness of one population is shown.

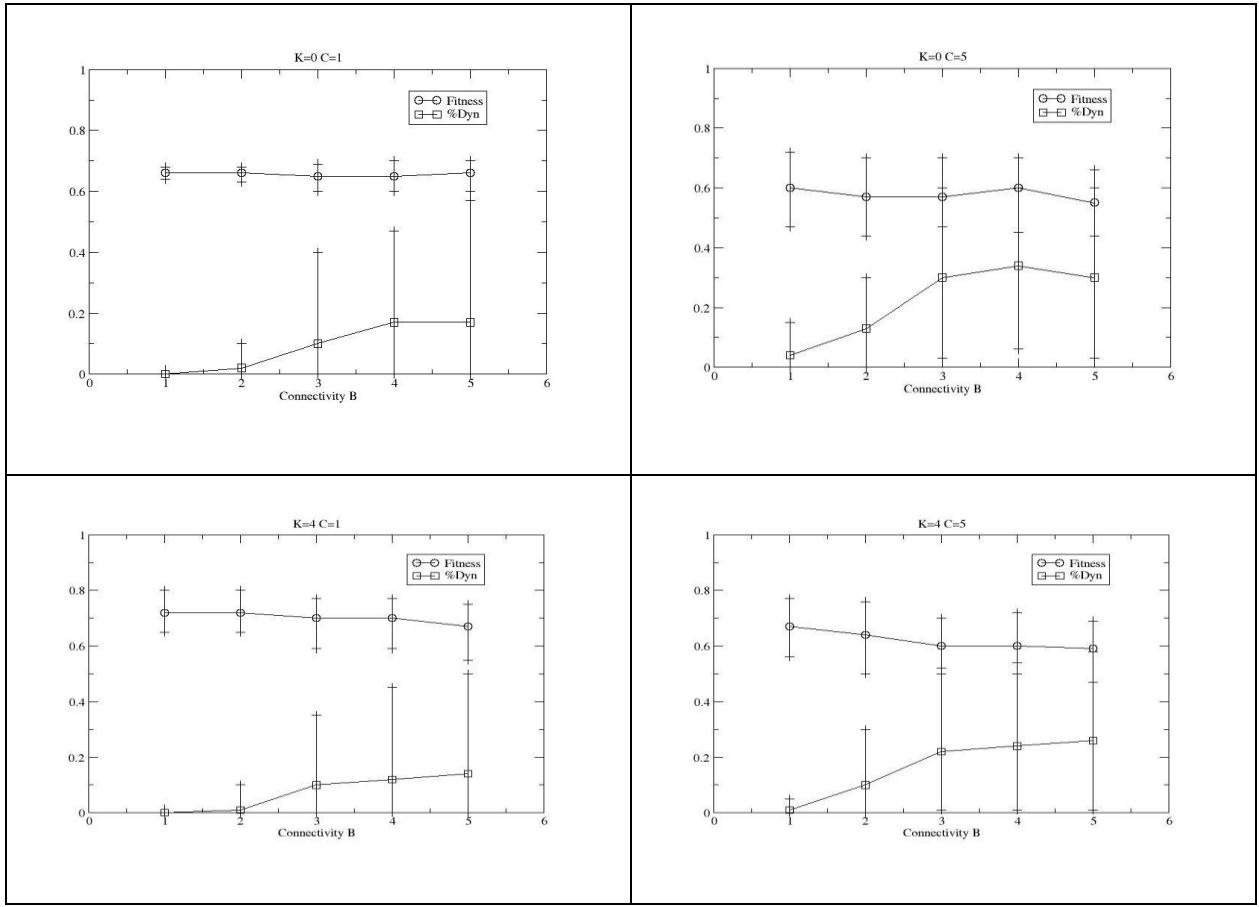

Fig 10: Performance of the fully dynamic RBN coevolved against another, after 30,000 generations.

Figure 10 shows that for *B*=1, regardless of *K*, dynamism is not selected for with low *C* but is selected for in around 2% of nodes on average for high *C*. For *B*=2 there is on average around 2% of dynamic nodes per GRN over various *K* for low *C* and 10% for high *C*. There is typically an increase in the amount of dynamism for the same *B* and *K* between *C*=1 and *C*=5 here, although with high variance. Again, there are roughly equal amounts of structural and functional dynamism. In these cases, unlike the multicellular case above, by constantly varying structure and function, the coupled networks are exhibiting uncooperative behaviour: the ability to dynamically alter the attractor space appears to make it easier for the coupled networks to compensate for or cause changes in behaviour. The ability of evolution to exploit dynamism in the high *B* cases is seen again, however this behaviour, unlike in the multicellular scenario, is not sensitive to *C*. The amount of dynamism rises to around 20% for low *C* and increases further for high *C*.

Figure 11 shows example single runs of these systems, indicating how the degree of dynamic nodes varies temporally depending upon the underlying coevolutionary behaviour: during periods of stasis dynamism is selected against but it rises dramatically during periods of re-adaptation to a change in the partner species. This automatic adjustment based upon evolutionary performance is somewhat reminiscent of that displayed by self-adaptive mutation rates in non-stationary environments (e.g., [Hoffmeister & Back, 1992]).

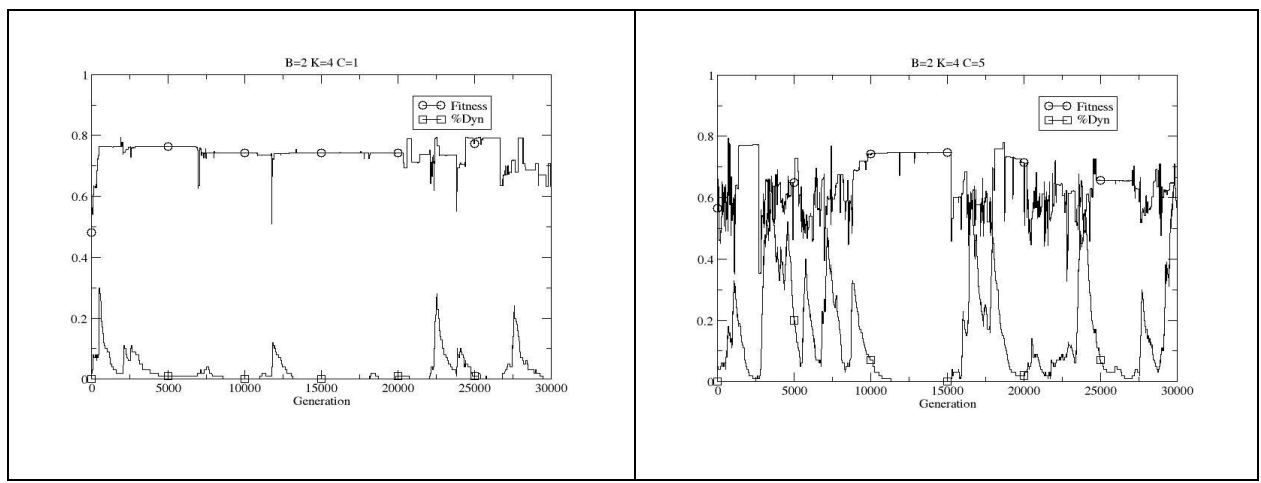

Fig 11: Example single runs of the coevolutionary case showing how dynamism is exploited to varying degrees depending upon the overall temporal dynamics of the ecosystem, rising during periods of re-adaptation and falling during periods of stasis.

## 6  Conclusions

There is a growing body of work within evolutionary computation which explores representations more closely analogous to those seen in nature, i.e., artificial genetic regulatory networks. Adoption of these relative generic representations creates the opportunity to exploit new search mechanisms based upon the discoveries of microbiology since the inception of evolutionary computation. That is, since the middle of the last century, molecular biologists have identified a variety of mechanisms through which changes in DNA sequences occur in natural regulatory networks in ways other than copy errors and recombinations: specific biochemical processes generate novelty through targeted DNA restructuring based upon the internal and external state of a GRN *during* the organismal lifecycle. This paper has explored the use of novel mechanisms based upon transposons within an extended version of a well-known abstract GRN model.

Building upon recent results, it has been shown that simple structural and functional dynamism is positively selected for in non-stationary environments, either in isolation or when both mechanisms are available simultaneously. Moreover, any genomic rearrangements occurring during a parent's lifecycle can be inherited by the offspring without detriment. The new mechanisms were also found to be selected for in coevolutionary environments, aiding evolutionary search by reducing sensitivity to the underlying dynamics of the GRN.

Further consideration of mobile DNA suggests a number of extensions to this work in the near future, particularly the use of retrotransposon-like copying to enable new mechanisms for changing the size of the GRN (see [Bull, 2012b]). Approaches drawing upon that recently presented by Smith et al. [2011] would seem appropriate, for example. Future work should also consider the use of transposon-inspired mechanisms within other GRN representations, explore whether these results also hold for asynchronous GRN updating schemes, determine whether the general results also appear to be true for much larger coevolutionary systems, i.e., for $S>1$, extend work using GRN as general purpose indirect encodings for other representations (after [Bull, 2012c]), etc.